\documentclass[showpacs,aps,twocolumn,superscriptaddress,prb,amsmath]{revtex4}
\usepackage{txfonts}
\usepackage{color}
\usepackage[latin9]{inputenc}
\usepackage{amssymb}
\usepackage{graphicx}% Include figure files
\usepackage{dcolumn}% Align table columns on decimal point
\usepackage{bm}% bold math
\usepackage{hyperref}
\usepackage{color}

\begin{document}
\title{Flat-band based ferromagnetic semiconducting state in the graphitic C$_4$N$_3$ monolayer}
\author{Chaoyu He}
\affiliation{Laboratory for Quantum Engineering and Micro-Nano Energy Technology and School of Physics and Optoelectronics, Xiangtan University, Hunan 411105, P. R. China}
\author{Yujie Liao}
\affiliation{Laboratory for Quantum Engineering and Micro-Nano Energy Technology and School of Physics and Optoelectronics, Xiangtan University, Hunan 411105, P. R. China}
\author{Tao Ouyang}
\email{ouyangtao@xtu.edu.cn}
\affiliation{Laboratory for Quantum Engineering and Micro-Nano Energy Technology and School of Physics and Optoelectronics, Xiangtan University, Hunan 411105, P. R. China}
\author{Huimin Zhang}
\email{hm$\_$zhang@fudan.edu.cn}
\affiliation{Key Laboratory of Computational Physical Sciences (Ministry of Education), State Key Laboratory of Surface Physics,
and Department of Physics, Fudan University, Shanghai 200433, China}
\author{Hongjun Xiang}
\email{hxiang@fudan.edu.cn}
\affiliation{Key Laboratory of Computational Physical Sciences (Ministry of Education), State Key Laboratory of Surface Physics,
and Department of Physics, Fudan University, Shanghai 200433, China}
\author{Jianxin Zhong}
\affiliation{Laboratory for Quantum Engineering and Micro-Nano Energy Technology and School of Physics and Optoelectronics, Xiangtan University, Hunan 411105, P. R. China}

\begin{abstract}
A new set of lattice-models based on the hexagonal $\sqrt{N}\times\sqrt{N}$ super-cells of the well-known honeycomb lattice with single-hole defect (HL-D-1/2N) are proposed to realize the nontrivial isolated flat-bands. Through performing both tight-binding and density functional theory calculations, we demonstrate that the experimentally realized graphitic carbon nitride (Adv. Mater., 22, 1004, 2010; Nat. Commun., 9, 3366, 2018), the HL-D-1/8 based C$_4$N$_3$, is a perfect system to host such flat bands. For the flat high-energy P-6m2 C$_4$N$_3$ structure, it displays the ferromagnetic half-metallicity which is not related to the isolated flat bands. However, the P-6m2 C$_4$N$_3$ structure is dynamically unstable. Using a structure searching method based on group and graph theory, we find that a new corrugated Pca21 C4N3 structure has the lowest energy among all known C$_4$N$_3$ structures. This Pca21 C$_4$N$_3$ structure is an intrinsic ferromagnetic half-semiconductor (Tc$\approx$241 K) with one semiconducting spin-channel (1.75 eV) and one insulating spin-channel (3.64 eV), which is quite rare in the two-dimensional (2D) systems. Its ferromagnetic semiconducting property originates from the isolated p$_z$-state flat-band as the corrugation shift the flat band upward to the Fermi level. Interestingly, this Pca21 C$_4$N$_3$ structure is found to be piezoelectric and ferroelectric, which makes C$_4$N$_3$ an unusual transition-metal-free 2D multiferroic.
\end{abstract}

\maketitle
%%%%%%%%%%%%%%%%%%%%%%%%%%%%%%%%
%\section{Introduction}
\indent The band structure of valence electrons affects most of the physical properties of condensed materials. Two limiting cases classified according to effective mass of carriers have attracted widely research interests, namely, the linear-dispersion band \cite{1} and the dispersionless flat-band \cite{3, 4, 5} holding massless Dirac fermion and infinitely heavy fermion, respectively. The linear-dispersion band has been experimentally detected in graphene monolayer \cite{1}, and has attracted numerous efforts to explore material candidates \cite{6, 7, 8} beyond graphene for realizing the fascinating physical phenomena of relativistic effect \cite{H1, H2} and Fractional Quantum Hall Effect \cite{R1, R2}. The dispersionless flat-band can also be realized in the graphene-based moire supper-lattice \cite{M1, M2}. It has been widely investigated in lattice models\cite{3, 4, 5, L3, L7, np2022} in the past decades and can also be theoretically realized in many stoichiometric materials \cite{3, 4, 5, LM1, LM3, LM6, LM10}. Exotic quantum phases, such as high temperature fractional quantum Hall effects \cite{HH1}, Bose-Einstein condensation \cite{BE}, Wigner crystallization \cite{WC1, WC2}, high-temperature superconductivity \cite{HS2} and ferromagnetism \cite{LM6, FM2}, are expected to be realized in flat-band systems. Regrettably, definite experimental observations of electronic flat-band in two-dimensional (2D) stoichiometric systems are still quite rare \cite{5, LM1}, which seriously hinders the realization of these exotic quantum states.

\indent Half-filled flat-bands usually induce ferromagnetism in order to lower the Hubbard repulsion \cite{mg1, mg2}. In particular, isolated flat-bands provides new possibilities to design 2D ferromagnetic (FM) semiconductors with a higher Curie temperature than known FM semiconductors (e.g., CrX$_3$ (X=Cl, Br, I)\cite{x8, x9}, VI$_3$ \cite{x11}, CrGeTe$_3$ \cite{x13} and CrSBr \cite{x14}) for applications in nanoscale spin-electronics \cite{APL-JX, NSR-YJL}, such as the single-spin transistor \cite{x1}, magnetic tunnel junctions \cite{x4} and some other spin-dependent devices \cite{x7, x5}. However, most of the previously proposed flat-bands are full-filled and they always touch together with a dispersive band with a wide band width larger than the spin-split energy, which prevents from the formation of FM semiconductor. For example, the one-hole doped kagome graphene (with kagome-type flat-band) is a FM half-metal after considering spin-split effect \cite{WC2}. The hole doped covalent-organic frameworks with Lieb lattice induced flat bands \cite{LM3, LM6} were reported as FM metals or half-metals. Although the partially hydrogenated graphene with inter-locking circles bonding-pattern is a semiconductor with isolated flat-band, it is nonmagnetic (NM) and will become a FM metal after hole-doping \cite{LM10} since the weak spin-split effect cannot overcome the width of the nearly flat-band. Because the isolated flat-band near the Fermi-level is extremely rare in both lattice models \cite{PRB2010, np2022, prb2021} and material platforms \cite{Nanoscale, prb2022}, the intrinsic half-filled and isolated flat-bands in realistic materials are still lacking for realizing stable FM orders holding semiconducting feature.

\indent In this letter, a new set of lattice-models are proposed through introducing single-hole defect in the hexagonal super-cells ($\sqrt{N}\times\sqrt{N}$, N=3, 4, 7, 9, 12, 13 and 16) of the well-known honeycomb lattice (HL-D-1/2N) for designing isolated flat-bands based on the p$_z$-state waves related tight-binding method (TB) \cite{prb2022}. In super-cells with index N is integer multiple of 3, the introduced single-hole defect will induce a zero energy flat-band (ZEFB) touching with the Dirac-cone at the $\Gamma$-point. In super-cells with index N beyond multiple of 3, the Dirac-cones are broken and introduced ZEFBs are always isolated on the Fermi-level. The DFT-based band structures of the HL-D-1/3 based C$_2$N$_3$ and the HL-D-1/8 based C$_4$N$_3$ confirm our TB results well. As an experimentally synthesized material \cite{AM2010, NC2018}, the HL-D-1/8 based C$_4$N$_3$ is finally confirmed to be an intrinsic FM semiconductor with corrugated ground state Pca21 configuration, but not a FM half-metal as reported before \cite{NC2018, PRL2012}. It is not only an excellent photocatalytic material \cite{r2, r6} but also an unusual metal-free 2D multiferroic with FM semiconducting, ferroelectric and piezoelectric properties for information application.
\begin{figure}
\begin{center}
\includegraphics[width=\columnwidth]{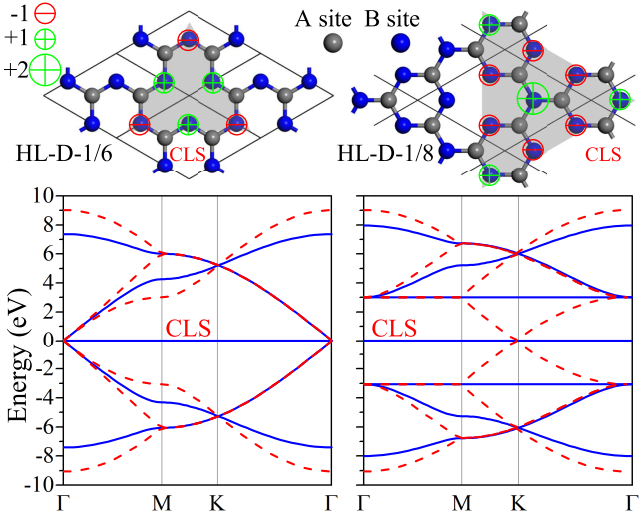}
\caption{The hexagonal super-cells ($\sqrt{N}\times\sqrt{N}$, N=3 and 4) of the well known honeycomb lattice (HL) with one-hole defect (HL-D-1/6 and HL-D-1/8) together with their corresponding P$_z$-wave based TB band structures (Blue solid lines). The band structures (red dash lines) of the perfect HL calculated in corresponding super-cells are also plotted to show the band folding effects.}
\end{center}
\end{figure}

\indent The hexagonal super-cell models named as HL-D-1/2N (N=3, 4, 7, 12, 13 and 16) are shown in Fig.S1 together with the corresponding perfect HL. The p$_z$-state electron waves in these lattice-models are simply described by the widely-used TB models as
\begin{equation}
 \begin{split}
&H=-\sum_{i\neq j}t_{ij}(c_{i}^{\dagger}c_{j}+h.c.)+\varepsilon\sum_{i}c_{i}^{\dagger}c_{i} \\
 \end{split}
\end{equation}
where $c_{i}^{\dagger}$ and $c_{j}$ are the creation and annihilation operators, respectively. The on-site energies $\varepsilon$ and the nearest hopping integrals t$_{ij}$ between the $i$-th and $j$-th sites can be considered as different parameters according to the symmetry of the lattice. In the perfect HL, A and B sublattices share same on-site energy and the adjacent hopping integrals between them are also equal to each other. It contains two dispersive p$_z$-wave bands with E$_1$(\textbf{k})=-E$_2$(\textbf{k}) for any \textbf{k} vector due to the chiral (sublattice) symmetry of SH(\textbf{k})S$^{-1}$=-H(\textbf{k}), where S is the Pauli matrix of $\sigma$$_3$. And these two bands degenerate at the high-symmetry K point of (1/3, 1/3, 0) forming the Dirac-cone feature due to the C$_3$ symmetry of this k-vector. The introduced single-hole defect (deleting one A site) does not broke the chiral symmetry of the super-cell and will introduce an exactly half-filled ZEFB on the Fermi-level due to the odd number (2N-1) of lattice sites (energy bands). As the results shown in Fig. S1 and Fig. 1 for super-cells with index N is integer multiple of 3, the two Dirac-cones in the perfect HL (unit cell) at the high symmetry K and K' points will be folded to $\Gamma$-point forming a four-fold degenerating point. The introduced single-hole defect break only one of the Dirac-cones and the induced ZEFB always touch the rested Dirac-cone at the $\Gamma$-point. In other super-cells with index N beyond multiple of 3, the two-fold degenerating Dirac-cones of K and K' cannot be folded to same k-point and they will be broken by the introduced single-hole defect leaving the corresponding ZEFB isolated. These ZEFBs can be understood as the destructive interference between the compact localized states (CLS) in the super-lattices as shown in Fig. 1 of HL-D-1/6 and HL-D-1/8 as example. And they must localize in the majority sublattices (namely the B sublattice in the super-lattices), similar to the case discussed previously \cite{prb2022}.

\indent When the discrepancies between lattice sites and linkers are considered, the combinations of on site and hopping energies are very complex. Here we just randomly sample 11 groups of symmetry-dependent parameters for the small-size models of HL-D-1/6 and HL-D-1/8 as the results shown in Fig. S2 and S3, respectively. Lattice HL-D-1/8 can be decomposed into three sublattices of Kagome A (KSL-A), Kagome B (KSL-B) and triangular B (TSL-B) as shown in Fig. 2. Thus, its band structure contains corresponding three band-groups including two kagome-type bands (KTBs: including a symmetry-protected flat-band touching with one of the two crossing dispersive bands) from KSL-A and KSL-B, as well as an isolated flat-band from TSL-B. The solitary of the isolated flat band is robust against the parameter variations but its flatness and relative positions are obviously affected by the used parameters. HL-D-1/6 consists of a coloring triangle B sublattice (CTSL-B) and a hexagonal A sublattice (HSL-A) as denoted in Fig. S4. Its band structure contains a KTBs from CTSL-B and a graphene-type bands (GTBs: two crossing dispersive bands) from HSL-A. The relative energy positions of KTBs and GTBs also depend on the used on site energies for CTSL-B (E$_B$) and HSL-A (E$_A$) as shown in Fig. S2.

\indent To realize these ZEFBs in realistic atomic systems, two imaginary carbon nitride phases with p$_z$-orbital are constructed based on HL-D-1/6 (C$_2$N$_3$) and HL-D-1/8 (C$_4$N$_3$) as shown in Fig. S4 and Fig.2, which satisfy the corresponding sp$^2$-hybridization features of 3-connected carbon and 2-connected nitrogen. As the results shown in Fig. S4 and Fig.2, the projected p$_z$-states of C$_2$N$_3$ and C$_4$N$_3$ are very similar to those calculated form TB method presented in Fig. 1. The "ZEFBs" are distorted due to the including of long-range interactions under first-principles calculations and they are always buried by the p$_x$ and p$_y$ states contributed from the 2-connected nitrogen atoms. When spin-split effect is considered, both the flat HL-D-1/6 based C$_2$N$_3$ and the flat HL-D-1/8 based C$_4$N$_3$ (Fig.4) will become FM half-metals.

\indent It should be noticed that the HL-D-1/8 based C$_4$N$_3$ has been experimentally synthesized before \cite{AM2010, NC2018} and it was reported as a half-metal in both exactly-flat and slightly-corrugated configurations \cite{PRL2012, NC2018}. As another HL-D-1/8 based graphitic carbon nitride, the 2D C$_3$N$_4$ monolayer prefers also a highly corrugated configuration as the ground state \cite{zln2021}. Thus, it is interesting to systematically search for the ground state configurations for the free-standing graphitic C$_4$N$_3$ monolayer and investigate the corrugation effects (crystal-field splitting) on the corresponding electronic and magnetic properties. The random method based on group and graph theory as implemented in our previously developed RG2 code \cite{Shi18, hcyprl} is employed to search the possible corrugated configurations for the free-standing 2D graphitic C$_4$N$_3$ monolayer. And the density functional theory (DFT) based first-principles method as implemented in the widely used VASP code \cite{VASP} is used, as described in part-I in the supplementary file, to optimize the structural candidates and identify the ground state configurations, as well as investigate the fundamental electronic, mechanical and magnetic properties of the low-energy configurations.
\begin{figure}
\begin{center}
\includegraphics[width=\columnwidth]{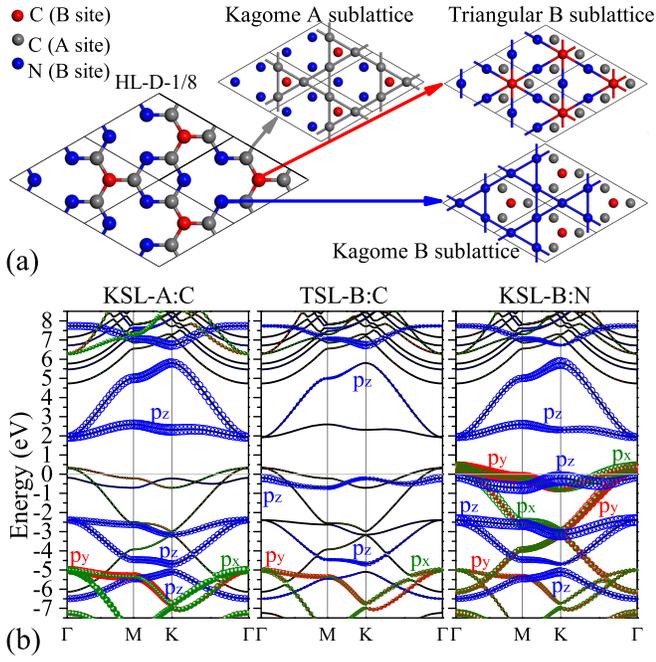}
\caption{(a) The crystal structure of the HL-D-1/8 bashed C$_4$N$_3$ monolayer in flat P-6m2 symmetry and the three corresponding sublattices of a carbon-filled Kagome A sublattice (KSL-A:C), a nitrogen-filled Kagome B sublattice (KSL-B:N) and a carbon-filed triangular B sublattice (TSL-B:C). (b) The projected p-state waves (p$_x$, p$_y$ and p$_z$ orbitals) of the KSL-A:C, KSL-B:N and TSL-B:C atoms on the calculated NM state band structure of the flat P-6m2 C$_4$N$_3$.}
\end{center}
\end{figure}

\indent As shown in Fig. S5 (a), the flat configuration with 7 atoms (HL-D-1/8 based C$_4$N$_3$) in P-6m2 symmetry is considered as the starting point. Nine inequivalent super-cells with atom number less than 28 (4 C$_4$N$_3$ units), including 3 hexagonal (H: $\times$1, $\times$3 and $\times$4), 3 rectangular (R: $\times$2, $\times$4 and $\times$4) and 3 monoclinic ones (M: $\times$2, $\times$3 and $\times$4), are constructed by RG2 for performing hypothetical (non-physical) phase transitions. The symmetry of each super-cell is randomly weaken into sub-groups to carry out position perturbation and structure optimization for sampling possible corrugated configurations. The quotient graph for each selected super-cell is kept to be (C: C C C), (C: C N N) and (N: C C) as that of the flat P-6m2 one in the whole process. We finally obtain 98 possible corrugated reconstructions for the free-standing graphitic C$_4$N$_3$ monolayer, including the previously proposed Pnc2 reconstruction \cite{PRL2012}. These corrugated configurations are systematically optimized by first principles calculations in both FM and NM states and their calculated total energies are plotted in Fig. S5 (b) as a function of the corresponding layer-thicknesses. Nearly all the corrugated candidates are more favorable than the flat P-6m2 one in energy and the FM states are always more stable than the NM states in any configurations. Especially, there are two new structural candidates (P321 and Pca21) with outstanding energetic stability exceeding than the previously proposed Pnc2 reconstruction as shown in Fig. 3. The structure relations and stabilities of the flat P-6m2, as well as the reconstructed Pnc2, P321 and Pca21 are discussed in part-II in the supplementary file as shown in Fig. S5 and S6.

\indent Low-energy means high probability of existence in nature or experimental conditions if the phase is dynamically stable against small lattice vibration. We have simulated the vibrational spectrums of the flat P-6m2 as well as the corrugated new ground state Pca21 in both FM and NM states to check their dynamical stabilities. Their calculated phonon band structures and phonon density of states are shown in Fig. S7 and Fig. S8, respectively. It is clearly that the flat P-6m2 one is dynamically unstable in both FM and NM situations, while the corrugated Pca21 is dynamically stable in FM state but unstable in NM situation. The calculated elastic constants C$_{11}$, C$_{12}$, C$_{22}$ and C$_{66}$ for Pca21 are 124.79 N/m, -20.38 N/m, 84.10 N/m and 70.37 N/m, respectively, which suggest that Pca21 is also mechanically stable to resist small-deformations according to the mechanical stability criteria (C$_{11}$$\times$C$_{22}$-C$_{12}^2$$>$0,C$_{66}$$>$0) for 2D materials.
\begin{figure}
\begin{center}
\includegraphics[width=\columnwidth]{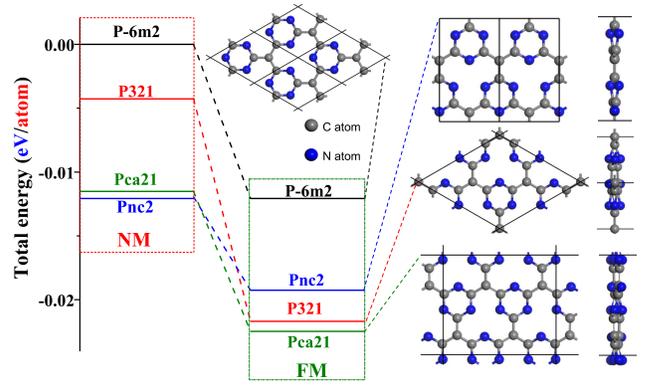}
\caption{The total energies of the flat P-6m2, the Pnc2 reconstruction, as well as the new ground state of Pca21 in NM and FM states. NM state energy of the flat P-6m2 one is selected as reference.}
\end{center}
\end{figure}

\indent Both flat P-6m2 and Pnc2 reconstruction are FM half-metals as their band structures shown in Fig. 4(a) and Fig. S9, which are in good agreement with the previous conclusions \cite{NC2018, PRL2012}. However, the corrugated ground state Pca21 is identified as a FM semiconductor as its band structure and density of state shown in Fig. 4(a). Intrinsic FM semiconductors are quite rare in 2D materials and we have carefully confirmed such an exciting discover through the high-level HSE06 method as shown in Fig. S10. In fact, the corrugated C$_4$N$_3$ in Pca21 symmetry can be classified as a FM half-semiconductor holding one semiconducting channel with band gaps of 1.748 eV and one insulating channel with larger band gaps of 3.641 eV. To evaluate its Curie temperature, a spin effective Hamiltonian can be constructed as
\begin{equation}
 \begin{split}
&H=\sum_{ij}\emph{J}_{ij}\emph{S}_{i}\cdot\emph{S}_{j} \\
 \end{split}
\end{equation}
under the nearest exchange interaction approximate, where \emph{S}$_i$ (\emph{S}$_j$) indicates the spin vector on site i (j) and \emph{J}$_{ij}$ is the exchange interaction between i and j. Since each localized state is centered at the carbon atom in TSL-B lattice, we regard the carbon atoms in TSL-B lattice as the spin sites, while the magnetic moments on nitrogen atoms are induced by C atoms in TSL-B lattice. Using the DFT total energies of two different spin configurations, the nearest spin exchange interactions (\emph{J}$_1$) can be deduced through energies mapping. In the C$_4$N$_3$ structure with Pca21 symmetry, \emph{J}$_1$ is evaluated to be -17.63 meV with setting total effective spin S=1. The corresponding Curie temperature of Pca21 is estimated to be 241 K based on the Monte Carlo simulation, which is much higher than those of the experimental synthesized 2D FM semiconductors.

\indent The origin of the FM semiconducting property in corrugated Pca21 can be revealed by comparing the NM band structure (Fig. 2) of the flat P-6m2 structure with that (Fig. S11) of the Pca21 structure. It is easy to see that the bands nearby the Fermi-level in these two systems are mainly contributed from the p$_z$ orbitals of KSL-A:C and TSL-B:C, as well as the p$_x$, p$_y$ and p$_z$ orbitals from KSL-B:N. We first discuss why flat P-6m2 is a half-metal as shown in Fig. 2, where it possesses two groups of p$_z$-state KTBs nearby the Fermi-level contributed from the KSL-A:C and KSL-B:N atoms and a full-filled p$_z$-state flat-band contributed mainly from the TSL-B:C atoms. Such a full-filled p$_z$-state flat-band is distorted to be nearly-flat-band and buried by the partially-filled p$_x$ and p$_y$ orbitals of the KSL-B:N atoms due to the including of the long-range interactions in first-principles calculations, which makes the flat P-6m2 a normal metal in NM situation. When the spin-degree is considered as shown in Fig. 4 (a), the p$_x$ and p$_y$ bands from the KSL-B:N atoms in the flat P-6m2 split into two subgroups of up and down, where the former is full-filled and the later is still partially-filled due to its wide dispersion-range. Thus, the flat P-6m2 becomes half-metal in its FM state. In this FM state, one can notice that the full-filled flat-bands from the p$_z$ orbitals from TSL-B:C are still full-filled in both spin-up and spin-down situations. Thus, the FM half-metallic property in the flat P-6m2 origin from the p$_x$ and p$_y$ based KTBs contributed from the nitrogen atoms in KSL-B, but not the isolated flat-band of the p$_z$ orbital from the carbon atoms in TSL-B, which can also be seen from the plot of the spin density (Fig. 4(a)).
\begin{figure}
\begin{center}
\includegraphics[width=\columnwidth]{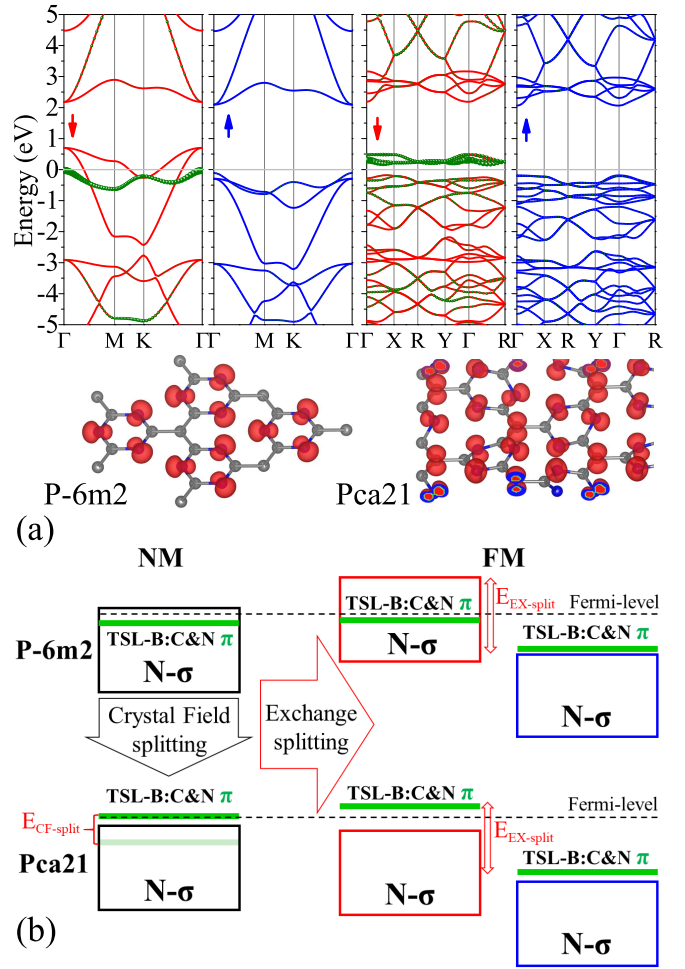}
\caption{The FM-state band structures with projected p$_z$ states (Olive solid balls) of the carbon atoms in TSL-B together with corresponding spin-charge densities (a) for the flat P-6m2 and the corrugated ground state Pca21. The orbital evolutions (b) from flat P-6m2 to corrugated Pca21 in both NM and FM states for showing split-effects of crystal-field (E$_{CF-split}$) and spin-exchange (E$_{EX-split}$).}
\end{center}
\end{figure}

\indent The schematic orbital-evolutions for only the p$_z$ orbitals from TSL-B:C and KSL-B:N atoms ($\pi$ band), as well as the p$_x$ and p$_y$ orbitals ($\sigma$ bands) from KSL-B:N near the Fermi-level are proposed in Fig. 4(b) to show the split-effects of crystal-field and spin-exchange. And the true evolutions can be noticed in Fig. S12, where the p$_x$ and p$_y$ bands from KSL-B:N and the isolated p$_z$-state flat-band from the TSL-B:C of Pca21 and P-6m2 (in compatible super-cell) are plotted together in both NM and FM states. With corrugation induced crystal-field splitting (E$_{CF-split}$), the $\pi$ band in Pca21 will be higher and thus isolated from the wide N-$\sigma$ bands with an obvious band gap of about 0.2 eV. This is because the corrugation weakens the electrostatic repulsion between the N $\sigma$ electrons. As shown in Fig. S12, the true $\pi$ bands from TSL-B:C and KSL-B:N in Pca21 form a narrow group locating in energy-range of about 0.42 eV and it is half-filled. When the spin-degree is considered, such a narrow group will further split into two sub-groups of spin-up and spin-down. In views of that the average split energy (E$_{EX-split}$) in Pca21 is obviously larger than the width of the narrow group, the sub-group of spin-up is exactly full-filled and the spin-down sub-group is totally unoccupied. Accordingly, the corrugated Pca21 becomes FM semiconductor due to the split effects of crystal-field and spin-exchange. The corrugation induced narrow band-groups with inter-group gaps are critical for forming the FM semiconducting properties in Pca21. The gradually opening of the inter-groups gaps as the increase of the layer-thickness of Pca21 are shown in Fig. S13 and the corresponding spin-split effects are shown in Fig. S14.

\indent The semiconducting property and the absence of inversion symmetry in C$_2v$ indicate that the corrugated Pca21 is a potential material with both piezoelectric and ferroelectric properties. There are three equivalent transition pathways from the flat P-6m2 to the corrugated Pca21 as shown in Fig. S15 and exist corresponding three equivalent ferroelectric phases (P1, P2 and P3) with in-plane spontaneous polarizations along three different directions. The three polarization directions form 120 degree angles with each other, which is different from the traditional case of two antiparallel polarization directions. Using the standard Berry phase method \cite{RMP}, the polarization in Pca21 C$_4$N$_3$ is calculated to be of 7.66$\times$10$^{-8}$ $\mu$C/cm which is equal to 5.86 $\mu$C/cm$^2$ in the three-dimensional situation when considering the monolayer thickness. The polarization is relatively larger than that in the 1T-MoS$_2$ (0.22 $\mu$C/cm$^2$) \cite{prl2014} and in the van der Waals interlayer-sliding ferroelectric BN (2.08$\times$10$^{-8}$ $\mu$C/cm in 2D and $\sim$0.68 $\mu$C/cm$^2$ in 3D) \cite{wmhacsnano}. The migration barrier between the two ferroelectric states is evaluated to be 0.29 eV/f.u. based on the climbing-image nudged elastic band (CI-NEB) method \cite{NEB}. The two piezoelectric coefficients of d11 and d12 for Pca21 are calculated to be 9.659 pm/V and 5.119 pm/V, respectively, which are even larger than that of MoS$_2$ monolayer \cite{piMoS2}.

\indent On the one hand, the isolated flat band causes ferromagnetism in graphitic C$_4$N$_3$ because of the corrugation of the structure. One the other hand, as the isolated flat band shifted upward by crystal-field splitting, there opens a band gap, which would reduce the screening by free carries and realize the ferroelectric polarization. It is worth to be noted that the graphitic C$_4$N$_3$ monolayer in corrugated Pca21 configuration is both ferromagnetic and ferroelectric. To our knowledge, it is the first prediction in flat band family that a material is reported to be multiferroic. This mechanism could supply a new way to design 2D multiferroics to searching the multiferroic materials in flat band system and broaden the research scope of flat band materials. Moreover, our study shows that such a small scale graphitic C$_4$N$_3$ monolayer with multiferroicity may have large potential application in the future high-density memories or electronics.

\indent In conclusion, a new set of lattice-models are proposed through introducing single-hole defect in the hexagonal super-cells ($\sqrt{N}\times\sqrt{N}$, N=3, 4, 7, 9, 12, 13 and 16) of the well-known honeycomb lattice (HL-D-1/2N) for designing isolated flat-band based on the p$_z$-state waves related tight-binding method (TB). We find that isolated flat bands can be realized in the super-cells with N is multiple of 3 under proper combinations of on-site and hopping intensities. The DFT-based band structures of two imaginary carbon nitride phases based on HL-D-1/6 (C$_2$N$_3$) and HL-D-1/8 (C$_4$N$_3$) well confirm the TB results. For the experimentally synthesized graphitic C$_4$N$_3$ monolayer as an isomorphic configuration with HL-D-1/8, its ground state is carefully confirmed as an intrinsic flat-band related FM semiconductor in corrugated Pca21 configuration but not the half-metallic flat P-6m2 or corrugated Pnc2. Pca21 is confirmed to be a FM half-semiconductor with one semiconducting spin-channel and one insulating spin-channel, which is quite rare in 2D materials. Its Curie temperature is evaluated to be 241 K, which is much higher than those of the experimental synthesized 2D FM semiconductors. Our discoveries highlight that the experimentally realized graphitic C$_4$N$_3$ is not only an excellent photocatalytic material in chemistry, but also an unusual metal-free 2D multiferroic with FM semiconducting, ferroelectric and piezoelectric properties for information application.

%\section{acknowledgement}
This work is supported by the National Natural Science Foundation of China (Grants No. 11974300 and 11974299), the Science Fund for Distinguished Young Scholars of Hunan Province of China (No. 2021JJ10036), and the Program for Changjiang Scholars and Innovative Research Team in University (No. IRT13093). Work at Fudan is supported by NSFC (Grants No. 11825403, No. 11991061, and No. 12188101) and Guangdong Major Project of Basic and Applied Basic Research (Future functional materials under extreme conditions-2021B0301030005).

%$\#$ These authors contributed equally to this work.
%%%%%%%%%%%%%%%%%%%%%%%%%%%%%%%%%%%%%%%%%%%%%%%%%%%%%%%%%%%%%%%%%%%%%
%% The same is true for Supporting Information, which should use the
%% suppinfo environment.
%%%%%%%%%%%%%%%%%%%%%%%%%%%%%%%%%%%%%%%%%%%%%%%%%%%%%%%%%%%%%%%%%%%%%
%\begin{suppinfo}

%\end{suppinfo}

%\subsection{References}
%+++++++++++++++++++++++++++++++++++++++++++++++++++++++++++++++++++++++++++++++++++++++++++++++++++++++++++++++++++++++++++++++++++++++++++++++++
\bibliographystyle{apsrev}

\end{document}